\documentclass[pra,12pt,tightenlines,showpacs,showkeys]{revtex4}
\usepackage{graphicx}
\usepackage{bbm}

\begin{document}

\title{Optimal Entanglement Witnesses for Qubits and Qutrits}
    \author{Reinhold A. Bertlmann}
    \author{Katharina Durstberger}
    \author{Beatrix C. Hiesmayr}
    \author{Philipp Krammer}
    \affiliation{Institute for Theoretical Physics, University of Vienna, Boltzmanngasse 5,
    A-1090 Vienna, Austria}

\begin{abstract}
We study the connection between the Hilbert-Schmidt measure of entanglement (that is the minimal
distance of an entangled state to the set of separable states) and entanglement witness in terms
of a generalized Bell inequality which distinguishes between entangled and separable states. A
method for checking the nearest separable state to a given entangled one is presented. We
illustrate the general results by considering isotropic states, in particular 2-qubit and 2-qutrit
states -- and their generalizations to arbitrary dimensions -- where we calculate the optimal
entanglement witnesses explicitly.
\end{abstract}

\keywords{entanglement, entanglement measure, entanglement witness, Hilbert-Schmidt distance, Bell
inequality, qutrit} \pacs{03.67.Mn, 03.67.Hk, 03.65.Ta, 03.65.Ca}

\maketitle

\section{Introduction}

Quantum entanglement is one of the most remarkable features of quantum mechanics
\cite{schroedinger35, einstein35}. In the last years it became clear that it can serve as a source
for various tasks in quantum information theory (see, e.g., Ref.~\cite{bertlmann02a}). Much
attention has been paid to explore the possibilities of applying quantum systems to communication
and computing protocols. Usually, these protocols use the information encoded in qubit systems;
however, higher dimensional systems, e.g. qutrits, are of increasing interest (see, e.g.,
\cite{brukner02}). Therefore it is important to get a more accurate description of entanglement,
especially for higher dimensional systems, which includes detecting and measuring entanglement
(for an overview see, e.g., Refs. \cite{bruss02, horodecki01}). For pure states such a description
is rather simple whereas for mixed states it is is more complicated.

The detection of entanglement, that is, distinguishing between separable and entangled states, has
become easy for 2-qubit states only. In this case necessary and sufficient conditions for
separability have been found \cite{peres96, horodecki96}, whereas for higher dimensions there
exist in general only necessary conditions for separability. In general, one can define several
types of entanglement measures, for instance, entanglement of formation \cite{bennett96}, the
concurrence \cite{hill97, wootters98} or the so called distance measures \cite{vedral97,
vedral98}.

In this paper a particular distance measure is used, the Hilbert-Schmidt distance, which
quantifies the distance of an entangled state to the set of all separable states. It is discussed
as an entanglement measure in Refs.~\cite{witte99, ozawa00}. We will call this measure shortly
Hilbert-Schmidt measure. In Ref.~\cite{bertlmann02} it is shown that the Hilbert-Schmidt measure
of an entangled state equals the maximal violation of the generalized Bell inequality which will
be discussed in this article.

The paper is organized as follows: In Sect.~\ref{secdef} we discuss the mathematical
basic concepts and definitions. In Sect.~\ref{secconnec} we re-examine shortly the
results of Ref.~\cite{bertlmann02} in order to get a lemma for determining the nearest
separable state to an entangled state. In Sect.~\ref{secqubitiso} and
Sect.~\ref{secqutritiso} we then illustrate our general results for the cases of
isotropic qubit and qutrit states. Finally, in Sect.~\ref{secqugeniso} we discuss
isotropic states in arbitrary dimensions.

\section{Concepts and Definitions} \label{secdef}

\subsection{Bipartite Systems in a Finite Dimensional Hilbert Space}

In this article we consider bipartite systems in a $d\times d$ dimensional Hilbert space
${\cal H}^d_A \otimes {\cal H}^d_B$. The observables acting in the subsystems ${\cal
H}_A$ and ${\cal H}_B$ are usually called Alice and Bob in quantum communication.
Observables are represented by Hermitian matrices and states by density matrices.

A state $\rho$ is called \emph{separable} if it can be written as a convex combination of
product states:
\begin{equation} \label{defsep}
    \rho_{\rm sep} \;=\; \sum_i p_i \, \rho^i_A \otimes \rho^i_B, \qquad 0 \leq p_i \leq 1, \
    \sum_i p_i = 1 \,.
\end{equation}
All states satisfying Eq. (\ref{defsep}) form the set of separable states $S$. If a state
is not separable, i.e., it cannot be written in terms of Eq.~(\ref{defsep}), then it is
called \emph{entangled}.

We define an \emph{isotropic} state $\rho_{\alpha}$ by (see Refs.
\cite{horodecki99, rains99, horodecki01})
\begin{equation} \label{rhodefiso}
    \rho_{\alpha} \;=\; \alpha \left| \phi_+^d \right\rangle \left\langle \phi_+^d \right|
    \,+\, \frac{1-\alpha}{d^2}\,\mathbbm{1}\,, \quad \alpha \in \mathbbm{R}\,, \quad
    - \frac{1}{d^2-1} \leq \alpha \leq 1 \;,
\end{equation}
where the range of $\alpha$ is determined by the positivity of the state. The state $\left|
\phi^d_+ \right\rangle$ is maximally entangled and given by
\begin{equation} \label{defiso}
    \left| \phi^d_+ \right\rangle \;=\; \frac{1}{\sqrt{d}}
    \sum_{i=0}^{d-1} \left| i \right\rangle_A \otimes \left| i \right\rangle_B\;,
\end{equation}
where $\left\{ \left| i \right\rangle \right\}$ is an orthonormal basis in ${\cal H}^d$.

The state is called isotropic because it is invariant under any $U_A \otimes U^*_B$
transformations (see Ref. \cite{horodecki99})
\begin{equation}
    (U_A \otimes U^*_B) \rho_{\alpha} (U_A \otimes U^*_B)^{\dag} \;=\; \rho_{\alpha} \,,
\end{equation}
where $U$ is a unitary operator, $U^*$ is its complex conjugate. The isotropic state
$\rho_{\alpha}$ has the following properties:
\begin{equation} \label{isosepent}
    \begin{array}{ccc}
        -\frac{\displaystyle 1}{\displaystyle d^2-1} \leq \alpha \leq
        \frac{\displaystyle 1}{\displaystyle d+1} &
        \quad\Rightarrow\quad & \rho_{\alpha} \;\mbox{ separable} \, , \\[2ex]
        \frac{\displaystyle 1}{\displaystyle d+1} < \alpha \leq 1 & \quad\Rightarrow\quad &
        \rho_{\alpha} \;\mbox{ entangled} \,.
    \end{array}
\end{equation}
Operators on a finite dimensional Hilbert space are elements of another Hilbert space
themselves, called Hilbert-Schmidt space ${\cal A} = {\cal A}_A \otimes {\cal A}_B$. In
this space the scalar product between two elements is defined as
\begin{equation}
     \left\langle A,B \right\rangle \;=\; \textnormal{Tr}\,A^{\dag}B \qquad A,B \in {\cal A} \,,
\end{equation}
with the corresponding Hilbert-Schmidt norm
\begin{equation} \label{defnorm}
    \left\|A\right\| \;=\; \sqrt{\left\langle A,A \right\rangle} \qquad A \in {\cal A} \, .
\end{equation}
\emph{Example for qubits.} In case of Alice and Bob acting on a
Hilbert space ${\cal H}^2_A \otimes {\cal H}^2_B$, an arbitrary
observable A can be written in the form
\begin{equation} \label{genobservable}
    A \;=\; a\,\mathbbm{1}_A \otimes \mathbbm{1}_B + a_i\,\sigma^i_A \otimes \mathbbm{1}_B +
    b_i\,\mathbbm{1}_A \otimes \sigma^i_B + c_{ij}\,\sigma^i_A \otimes \sigma^j_B, \quad\;
    a,a_i,b_i,c_{ij} \in \mathbbm{R} \,.
\end{equation}
Note that $c_{ij}$ can be diagonalized by 2 independent orthogonal transformations on
$\sigma^i_A$ and $\sigma^j_B$ \cite{henley62}. The operator A represents a density matrix
if $a = 1/4$ and $\sum_i ( a_i^2
+ b_i^2 ) + \sum_{i,j} c_{ij}^2 \leq 1/16\,$.\\

With help of the norm (\ref{defnorm}) we can quantify a distance between two arbitrary
states $\rho_1,\rho_2$, the \emph{Hilbert-Schmidt distance},
\begin{equation} \label{defhsdistance}
    d_{\rm HS}(\rho_1,\rho_2) \;=\; \left\| \rho_1 - \rho_2 \right\| \,.
\end{equation}
Viewing the Hilbert-Schmidt distance as an entanglement measure (see Refs. \cite{witte99,
ozawa00}) we define the so-called \emph{Hilbert-Schmidt measure}
\begin{equation} \label{defhsmeasure}
    D(\rho_{\rm ent}) \;=\; \min_{\rho \in S} d_{\rm HS}(\rho,\rho_{\rm ent}) \;=\;
    \min_{\rho \in S} \left\| \rho - \rho_{\rm ent} \right\| \,,
\end{equation}
which is the minimal distance of an entangled state $\rho_{\rm ent}$ to the set of
separable states.\\

An \emph{entanglement witness} $A \in {\cal A}$ is a Hermitian operator that `detects'
the entanglement of a state $\rho_{\rm ent}$ via inequalities \cite{horodecki96,
terhal00, terhal02, bertlmann02}
\begin{eqnarray} \label{defentwit}
    \left\langle \rho_{\rm ent},A \right\rangle = \textnormal{Tr}\, \rho_{\rm ent} A & < & 0 \,,
    \nonumber\\
    \left\langle \rho,A \right\rangle = \textnormal{Tr}\, \rho A & \geq & 0 \qquad
    \forall \rho \in S \,.
\end{eqnarray}

\emph{Geometric illustration.} For a geometrical illustration of the above inequalities let us
consider the following:
\begin{figure}
    \centering
        \includegraphics[width=0.50\textwidth]{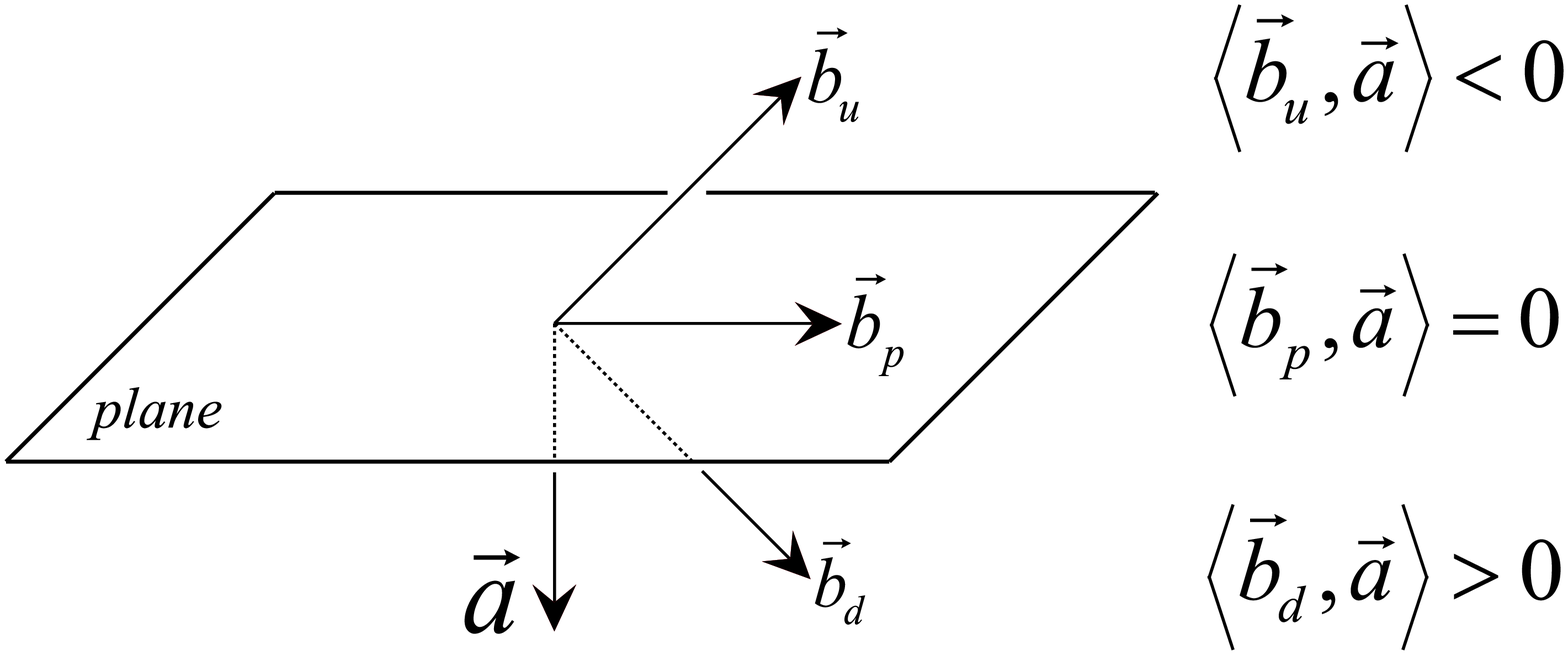}
    \caption{Geometric illustration of a plane in Euclidean space and the different values of
    the scalar product for states above ($\vec{b}_u$), within ($\vec{b}_p$) and
    below ($\vec{b}_d$) the plane.}
    \label{figeucplane}
\end{figure}
In Euclidean space a plane is defined by its orthogonal vector $\vec{a}$. The plane
separates vectors which have a negative scalar product with $\vec{a}$ from vectors having
a positive one; vectors in the plane have, of course, a vanishing scalar product (see
Fig.~\ref{figeucplane}).

This can be compared with our situation: A scalar functional $\left\langle \rho,A
\right\rangle = 0$ defines a hyperplane in the set of all states, and this plane
separates `left-hand' states $\rho_l$ satisfying $\left\langle \rho_l , A \right\rangle
<0$ from `right-hand' states $\rho_r$ with $\left\langle \rho_r , A \right\rangle >0\,$.
States $\rho_p$ with $\left\langle \rho_p , A \right\rangle = 0$ are inside the
hyperplane. According to the Hahn-Banach theorem, one can conclude that due to the
convexity of the set of separable states, there always exists a plane
that separates an entangled state from the set of separable states.\\

\begin{figure}
    \centering
        \includegraphics[width=0.40\textwidth]{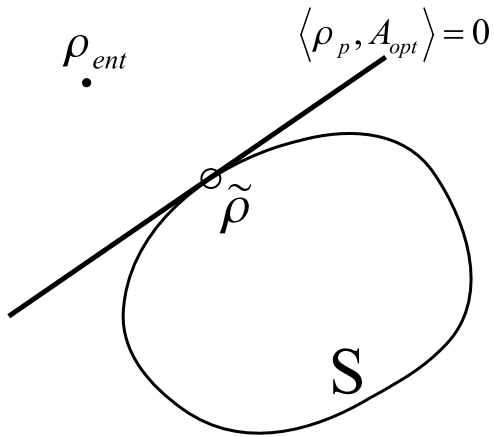}
    \caption{Illustration of an optimal entanglement witness}
    \label{figentopt}
\end{figure}

An entanglement witness is `optimal', denoted by $A_{opt}$, if apart from
Eq.~(\ref{defentwit}) there exists a separable state $\tilde{\rho} \in S$ such that
\begin{equation}
    \left\langle \tilde{\rho},A_{opt} \right\rangle \;=\; 0 \,.
\end{equation}
It is optimal in the sense that it defines a tangent plane to the set of separable states
$S$ and is therefore called \emph{tangent functional} \cite{bertlmann02}; see
Fig.~\ref{figentopt}.
\\

According to Ref.~\cite{bertlmann02}, we call the lower one of the inequalities (\ref{defentwit})
a \textit{generalized Bell inequality}, short GBI. `Generalized' means that it detects
entanglement and not just non-locality. Thus it doesn't serve as a criterion to distinguish
between a local hidden variable (LHV) theory and quantum theory as the usual Bell inequality does.
However, pay attention that in literature the term `generalized Bell inequalities' is also often
used for inequalities that detect non-locality, but are of more general form (more measurements,
etc.) than Bell's original inequality (see, e.g., Refs. \cite{terhal00, bertlmann01}). Bell
inequalities, like the CHSH inequality (Clauser, Horne, Shimony, Holt) \cite{clauser69}
\begin{equation} \label{chsh}
    \left\langle \rho , 2 \mathbbm{1} - B \right\rangle \;\geq\; 0 \;, \qquad B \;=\;
    \vec{a} \cdot \vec{\sigma} \otimes (\vec{b} + \vec{b}') \cdot \vec{\sigma} +
    \vec{a}' \cdot \vec {\sigma} \otimes (\vec{b} - \vec{b}') \cdot \vec{\sigma} \,,
\end{equation}
with unit vectors $\vec{a},\vec{a}',\vec{b},\vec{b}'\in \mathbbm{R}^3$ do \emph{not} necessarily
detect entanglement. But the inequality (\ref{chsh}) has to be satisfied by any state $\rho$ that
admits a LHV model. There exist examples of entangled states -- so-called Werner states
\cite{werner89} -- that do not violate the CHSH inequality. Nevertheless, \emph{every} entangled
state violates the GBI for an appropriate entanglement witness $A$.\\

Let us re-write Eq. (\ref{defentwit}) as
\begin{equation} \label{gbi}
    \left\langle \rho,A \right\rangle \,-\, \left\langle \rho_{\rm ent},A \right\rangle \;\geq\; 0
    \qquad \forall \rho \in S \,.
\end{equation}
The \emph{maximal} violation of the GBI is defined by
\begin{equation} \label{maxviolationgbi}
    B(\rho_{\rm ent}) = \max_{A, \, \left\| A - a \mathbbm{1} \right\| \leq 1} \left( \min_{\rho \in S}
    \left\langle \rho,A \right\rangle - \left\langle \rho_{\rm ent},A \right\rangle \right),
\end{equation}
where the maximum is taken over all possible entanglement witnesses $A$, suitably normalized, and
$a$ is the coefficient of the unity term of the general expression (\ref{genobservable}). A
general expression for quantifying entanglement with entanglement witnesses can be found in Ref.
\cite{Brandao}.

\subsection{Qubits} \label{basicsqubits}

A qubit state $\omega$, acting on ${\cal H}^2$, can be decomposed into Pauli matrices
\begin{equation} \label{qubitpauli}
    \omega \;=\; \frac{1}{2} \left( \mathbbm{1} + n_i\, \sigma^i \right), \qquad
    n_i \in \mathbbm{R}\,, \; \sum_i n_i^2 = \left| \vec{n} \right|^2 \leq 1 \,.
\end{equation}
Note that for $\left|\vec{n}\right|^2 < 1$ the state is mixed (corresponding to
Tr$\,\omega^2 < 1$) whereas for $\left|\vec{n}\right|^2 = 1$ the state is pure
(Tr$\,\omega^2 = 1$).

We can write any density matrix of 2-qubits $\rho$ acting on ${\cal H}^2 \otimes {\cal
H}^2$ (for convenience we drop the indices $A$ and $B$ from now on) in a basis of $4
\times 4$ matrices, the tensor products of the identity matrix $\mathbbm{1}$ and the
Pauli matrices $\sigma^i$,
\begin{equation} \label{statewithpauli}
    \rho \;=\; \frac{1}{4} \left( \mathbbm{1} \otimes \mathbbm{1} \,+\,
    a_i\,\sigma^i \otimes \mathbbm{1} \,+\, b_i\,\mathbbm{1} \otimes \sigma^i \,+\,
    c_{ij}\,\sigma^i \otimes \sigma^j \right)\,, \qquad
    a_i,b_i,c_{ij} \in \mathbbm{R} \,.
\end{equation}
A product state $\omega \otimes \rho$ has the form
\begin{eqnarray} \label{productpauli}
     \omega \otimes \rho &\;=\;& \frac{1}{4} \left( \mathbbm{1} \otimes  \mathbbm{1} \,+\,
    n_i\,\sigma^i \otimes \mathbbm{1} \,+\, m_i\,\mathbbm{1} \otimes \sigma^i \,+\,
    n_i m_j\, \sigma^i \otimes \sigma^j \right)\,,  \nonumber\\
    & & \quad n_i, m_i \in \mathbbm{R}\,, \; \left| \vec{n} \right| \leq 1\,, \;
    \left| \vec{m} \right| \leq 1 \,.
\end{eqnarray}
Any separable state can be written as the convex combination of
expression (\ref{productpauli}),
\begin{eqnarray} \label{seppauli}
     \rho_{\rm sep} \;=\; & \sum_k p_k & \, \frac{1}{4} \left( \mathbbm{1} \otimes \mathbbm{1}
    \,+\, n_i^k\,\sigma^i \otimes \mathbbm{1} \,+\, m_i^k\,\mathbbm{1} \otimes \sigma^i
    \,+\, n_i^k m_j^k \,\sigma^i \otimes \sigma^j \right)\,, \nonumber\\
    & & \quad\; n_i^k, m_i^k \in \mathbbm{R}\,, \; \left| \vec{n}^k \right| \leq 1\,, \;
    \left| \vec{m}^k \right| \leq 1 \,.
\end{eqnarray}

\subsection{Qutrits}

The description of qutrits is very similar to the one for qubits. A qutrit state $\omega$ on
${\cal H}^3$ can be expressed in the matrix basis $\left\{ \mathbbm{1}, \lambda^1, \lambda^2, \
\ldots \ ,\lambda^8 \right\}$ with an appropriate set $\{n_i\}$
\begin{equation}
    \omega \;=\; \frac{1}{3} \left( \mathbbm{1} + \sqrt{3} \,n_i \,\lambda^i \right), \qquad
    n_i \in \mathbbm{R}\,, \; \sum_i n_i^2 = \left| \vec{n} \right|^2 \leq 1 \;.
\end{equation}
The factor $\sqrt{3}$ is included for a proper normalization (see, e.g., Refs.~\cite{arvind97,
caves00}). The matrices $\lambda^i$ ($i=1,...,8$) are the eight Gell-Mann matrices
\begin{eqnarray} \label{defgellmann}
    & \lambda^1 = \left(
    \begin{array}{ccc}
        0 & 1 & 0 \\
        1 & 0 & 0 \\
        0 & 0 & 0 \\
    \end{array} \right), \quad
    \lambda^2 = \left(
    \begin{array}{ccc}
        0 & -i & 0 \\
        i & 0 & 0 \\
        0 & 0 & 0 \\
    \end{array} \right), \quad
    \lambda^3 = \left(
    \begin{array}{ccc}
        1 & 0 & 0 \\
        0 & -1 & 0 \\
        0 & 0 & 0 \\
    \end{array} \right), &
    \nonumber \\
    & \lambda^4 = \left(
    \begin{array}{ccc}
        0 & 0 & 1 \\
        0 & 0 & 0 \\
        1 & 0 & 0 \\
    \end{array} \right), \quad
    \lambda^5 = \left(
    \begin{array}{ccc}
        0 & 0 & -i \\
        0 & 0 & 0 \\
        i & 0 & 0 \\
    \end{array} \right), \quad
    \lambda^6 = \left(
    \begin{array}{ccc}
        0 & 0 & 0 \\
        0 & 0 & 1 \\
        0 & 1 & 0 \\
    \end{array} \right), &
    \nonumber \\
    &
    \lambda^7 = \left(
    \begin{array}{ccc}
        0 & 0 & 0 \\
        0 & 0 & -i \\
        0 & i & 0 \\
    \end{array} \right), \quad
    \lambda^8 = \frac{1}{\sqrt{3}} \left(
    \begin{array}{ccc}
        1 & 0 & 0 \\
        0 & 1 & 0 \\
        0 & 0 & -2 \\
    \end{array} \right), &
\end{eqnarray}
with properties $\mbox{Tr}\,\lambda^i = 0, \; \mbox{Tr}\,\lambda^i
\lambda^j = 2\, \delta^{ij}$.

A 2-qutrit state, acting on ${\cal H}^3 \otimes {\cal H}^3$, can be represented in a
basis of $9 \times 9$ matrices consisting of the unit matrix $\mathbbm{1}$ and the eight
Gell-Mann matrices $\lambda^i$
\begin{equation} \label{statewithgellmann}
    \rho \;=\; \frac{1}{9} \left( \mathbbm{1} \otimes \mathbbm{1} \,+\,
    a_i\,\lambda^i \otimes\mathbbm{1} \,+\, b_i\,\mathbbm{1} \otimes \lambda^i \,+\,
    c_{ij}\,\lambda^i \otimes \lambda^j \right), \qquad a_i,b_i,c_{ij} \in \mathbbm{R} \,.
\end{equation}
By the same argumentation as for qubits any separable 2-qutrit state is a convex
combination of product states
\begin{equation} \label{sepgellmann}
    \rho_{\rm sep} \;=\; \sum_k p_k \ \frac{1}{9} \left( \mathbbm{1} \otimes \mathbbm{1}
    \,+\, \sqrt{3}\,n_i^k\,\lambda^i \otimes \mathbbm{1}
    \,+\, \sqrt{3}\, m_i^k\,\mathbbm{1} \otimes \lambda^i
    \,+\, 3 \,n_i^k m_j^k \,\lambda^i \otimes \lambda^j \right) \;.
\end{equation}

\section{Connection Between Hilbert-Schmidt Measure and Entanglement Witness} \label{secconnec}

\subsection{Geometrical Considerations about the Hilbert-Schmidt Distance} \label{secgeoconhs}

Before we are going to discuss the Bertlmann-Narnhofer-Thirring Theorem \cite{bertlmann02} let us
consider the Hilbert-Schmidt distance. The geometrical illustration we are going to derive turns
out to be helpful for the proof of the Theorem.

We can write the Hilbert-Schmidt distance of any two states $\rho_1,\rho_2 \in{\cal A}$
as
\begin{equation}
    d_{\rm HS}(\rho_1,\rho_2) \;=\; \left\| \rho_1 - \rho_2 \right\| \;=\;
    \left\langle \rho_1 - \rho_2 ,
    \frac{\rho_1 - \rho_2}{\left\| \rho_1 - \rho_2 \right\|} \right\rangle \;=\;
    \left\langle \rho_1 - \rho_2 , \bar{C} \right\rangle \;.
\end{equation}
where we define the operator
\begin{equation}
    \bar{C} \;:=\; \frac{\rho_1 - \rho_2}{\left\| \rho_1 - \rho_2 \right\|} \;.
\end{equation}
Instead of $\bar{C}$ we may also choose $C := \bar{C} +
c\,\mathbbm{1}$ $(c \in \mathbbm{C})$ and find
\begin{equation}\label{generalhsconnection}
    d_{\rm HS}(\rho_1,\rho_2)=\left\langle \rho_1 - \rho_2 , \bar{C} \right\rangle\;=\;
    \left\langle \rho_1 - \rho_2 , \bar{C} \right\rangle +
    \left\langle \rho_1    - \rho_2 , c\,\mathbbm{1} \right\rangle \;=\;
    \left\langle \rho_1 - \rho_2 , C \right\rangle,
\end{equation}
since $\left\langle \rho_1 - \rho_2 , \mathbbm{1} \right\rangle =
\textnormal{Tr}\rho_1 - \textnormal{Tr}\rho_2 = 0\,$. For
convenience we fix $c$ to
\begin{equation} \label{defconstantc}
    c \;=\; - \frac{\left\langle \rho_1 , \rho_1 - \rho_2 \right\rangle}{\left\| \rho_1 -
    \rho_2 \right\|} \;,
\end{equation}
and obtain
\begin{equation} \label{defc}
    C \;=\; \frac{\rho_1 - \rho_2 - \left\langle \rho_1 ,
    \rho_1 - \rho_2 \right\rangle \mathbbm{1}}{\left\| \rho_1 - \rho_2 \right\|} \;.
\end{equation}

Analogously to Euclidean space we define a hyperplane $P$ that includes $\rho_1$ and is
orthogonal to $\rho_1 - \rho_2$ as the set of all states $\rho_p$ satisfying
\begin{equation} \label{defplane}
    \frac{1} {\left\| \rho_1 - \rho_2 \right\|} \left\langle \rho_p - \rho_1 ,
    \rho_1 - \rho_2 \right\rangle \;=\; 0 \;.
\end{equation}
For all states on one side of the plane, let us call them `left-hand' states $\rho_l$, we
have
\begin{figure}
    \centering
        \includegraphics[width=0.40\textwidth]{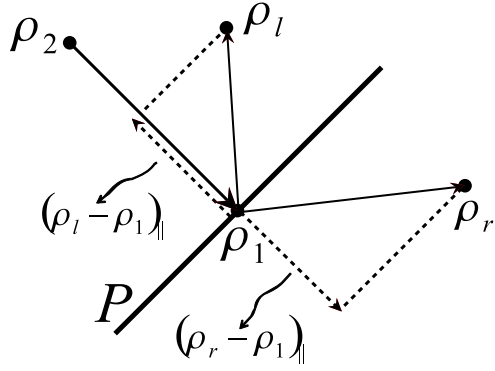}
    \caption{Illustration of Eqs. (\ref{left}) and (\ref{right}): The scalar product
    $\left\langle \rho_l - \rho_1 , \rho_1 - \rho_2 \right\rangle$ is \emph{negative} because
    the projection $\left( \rho_l - \rho_1 \right)_{\parallel}$ onto $\rho_1 - \rho_2$ points
    in the opposite direction to $\rho_1 - \rho_2$. On the other side,
    $\left\langle \rho_r - \rho_1 , \rho_1 - \rho_2 \right\rangle$ is \emph{positive} for
    states $\rho_r$, because then the projection $\left( \rho_r - \rho_1 \right)_{\parallel}$
    points in the same direction as $\rho_1 - \rho_2$.}
    \label{figproofhsg}
\end{figure}
\begin{equation} \label{left}
    \frac{1}{\left\| \rho_1 - \rho_2 \right\|} \left\langle \rho_l - \rho_1,
    \rho_1 - \rho_2 \right\rangle \;<\; 0 \;,
\end{equation}
whereas the states on the other side, the `right-hand' states $\rho_r$ are given by
\begin{equation} \label{right}
    \frac{1}{\left\| \rho_1 - \rho_2 \right\|} \left\langle \rho_r - \rho_1,
    \rho_1 - \rho_2 \right\rangle \;>\; 0 \;.
\end{equation}
For an illustration see Fig.~\ref{figproofhsg}.\\

We can re-write Eqs. (\ref{defplane}), (\ref{left}), and (\ref{right}) with help of
operator $C$ by using
\begin{eqnarray}
    \left\langle \rho\, , C \right\rangle & \;=\; & \left\langle \rho\, ,
    \frac{\rho_1 - \rho_2}{\left\| \rho_1 - \rho_2
        \right\|} \right\rangle \;-\; \frac{\left\langle \rho_1 ,
        \rho_1 - \rho_2 \right\rangle} {\left\| \rho_1 - \rho_2
        \right\|} \, \left\langle    \rho\, , \mathbbm{1} \right\rangle \nonumber\\
    & \;=\; & \frac{1}{\left\| \rho_1 - \rho_2 \right\|} \left\langle \rho - \rho_1,
        \rho_1 - \rho_2 \right\rangle \;.
\end{eqnarray}
Then the plane $P$ is determined by
\begin{equation} \label{defplanewithc}
    \left\langle \rho_p\, , C \right\rangle \;=\; 0\;,
\end{equation}
and the `left-hand' and `right-hand' states satisfy the inequalities
\begin{eqnarray} \label{leftright}
    \left\langle \rho_l\, , C \right\rangle \;<\; 0 \quad \mbox{ and} \quad
    \left\langle \rho_r\, , C \right\rangle \;>\; 0 \;.
\end{eqnarray}

\subsection{The Bertlmann-Narnhofer-Thirring Theorem} \label{sectheorem}

Interestingly, one can find connections between the Hilbert-Schmidt measure and the
concept of entanglement witnesses. In particular, there exists the following equivalence
stated in the Bertlmann-Narnhofer-Thirring Theorem \cite{bertlmann02}:
\begin{quote}
    \textbf{Theorem.} The Hilbert-Schmidt measure of an entangled state equals the maximal
    violation of the GBI:
\begin{equation} \label{bnttheorem}
    D(\rho_{\rm ent}) \;=\; B(\rho_{\rm ent}) \;.
\end{equation}
\end{quote}
\emph{Proof.} We want to prove the Theorem in a different way as in
Ref. \cite{bertlmann02}.

\begin{figure}
    \centering
        \includegraphics[width=0.40\textwidth]{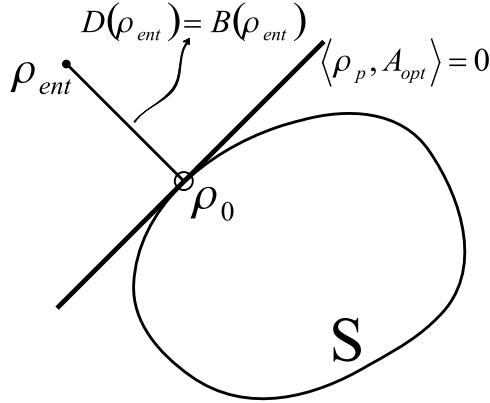}
    \caption{Illustration of the Bertlmann-Narnhofer-Thirring Theorem}
    \label{figbnttheo}
\end{figure}

For an entangled state $\rho_{\rm ent}$ the minimum of the Hilbert-Schmidt distance --
the Hilbert-Schmidt measure -- is attained for some state $\rho_0$  since the norm is
continuous and the set $S$ is compact
\begin{equation}
    \min_{\rho \in S} \left\| \rho - \rho_{\rm ent} \right\| \;=\;
    \left\| \rho_0 - \rho_{\rm ent} \right\| \,.
\end{equation}
In Eqs. (\ref{generalhsconnection}) and (\ref{defc}) we identify $\rho_1 = \rho_0$ and
$\rho_2 = \rho_{\rm ent}$ and with $C$ given by Eq. (\ref{defc}) we obtain the
Hilbert-Schmidt measure
\begin{equation} \label{specialhsconnection}
    d_{\rm HS}(\rho_0,\rho_{\rm ent}) \;=\; D(\rho_{\rm ent}) \;=\; \left\langle \rho_0 ,
    C \right\rangle - \left\langle \rho_{\rm ent} , C \right\rangle\, .
\end{equation}
In Eq.~(\ref{specialhsconnection}) the operator $C$ has to be an optimal entanglement
witness for the following reason: The state $\rho_0$ lies on the boundary of the set of
all separable states $S$ and the hyperplane defined by $\left\langle \rho_p\, , C
\right\rangle = 0$ is orthogonal to $\rho_0 - \rho_{\rm ent}$. Because $\rho_0$ is the
nearest separable state to $\rho_{\rm ent}$ the plane has to be tangent to the set $S$
(see Fig.~\ref{figbnttheo}). Eqs. (\ref{defplanewithc}), (\ref{leftright}) imply the
inequalities (\ref{defentwit}), it therefore follows that $C$ is an optimal entanglement
witness
\begin{equation} \label{entwitmaxviolation}
    A_{opt} \;=\; C \;=\; \frac{\rho_0 - \rho_{\rm ent} - \left\langle \rho_0 ,
    \rho_0 - \rho_{\rm ent} \right\rangle \mathbbm{1}}{\left\| \rho_0 - \rho_{\rm ent}
    \right\|}\;,
\end{equation}
which we use to rewrite the Hilbert-Schmidt measure (\ref{specialhsconnection})
\begin{equation}
    D(\rho_{\rm ent}) \;=\; \left\langle \rho_0 , A_{opt} \right\rangle -
    \left\langle \rho_{\rm ent} , A_{opt} \right\rangle \;.
\end{equation}
Note that in general the operator $C$ of Eq. (\ref{defc}) (where $\rho_1$ and $\rho_2$ are
arbitrary states) is not yet an entanglement witness.

Since the entanglement witness is optimal, i.e.,
\begin{equation}
\max_A \left( - \left\langle \rho_{\rm ent} , A \right\rangle \right) \;=\; - \left\langle
\rho_{\rm ent} , A_{opt} \right\rangle \;,
\end{equation}
where $A$ is restricted by $\left\| A - a \mathbbm{1} \right\| \leq 1$ and $\left\langle \rho_0 ,
A_{opt} \right\rangle = 0\,$, we obtain
\begin{eqnarray}
    D(\rho_{\rm ent}) & \;=\; & \left\langle \rho_0 , A_{opt} \right\rangle -
    \left\langle \rho_{\rm ent} , A_{opt} \right\rangle \;=\;
    \max_{A, \, \left\| A - a \mathbbm{1} \right\| \leq 1}
    \left( \left\langle \rho_0 , A \right\rangle -
    \left\langle \rho_{\rm ent} , A \right\rangle \right) \nonumber\\
    & \;=\; & \max_{A, \, \left\| A - a \mathbbm{1} \right\| \leq 1} \left(\min_{\rho \in S}
    \left\langle \rho , A \right\rangle - \left\langle \rho_{\rm ent} , A \right\rangle \right)
    \;=\; B(\rho_{\rm ent})\;,
\end{eqnarray}
which completes the proof.

Similar methods for constructing an entanglement witness can be found in Ref.~\cite{pittenger03};
for other approaches see, e.g., Refs.~\cite{lewenstein00, guehne03, ioannou04}.

\subsection{How to Check a Guess of the Nearest Separable State} \label{secguess}

Given an entangled state $\rho_{\rm ent}$, for the Hilbert-Schmidt measure we have to
calculate the minimal distance to the set of separable states $S$,
Eq.~(\ref{defhsmeasure}). In general it is not easy to find the correct state $\rho_0$
which minimizes the distance (for specific procedures, see, e.g., Refs.
\cite{verstraete01, zyczkowski98, zyczkowski99}). However, we can use an operator like in
Eq.~(\ref{defc}) for checking a good guess for $\rho_0$.

\begin{figure}
    \centering
        \includegraphics[width=0.40\textwidth]{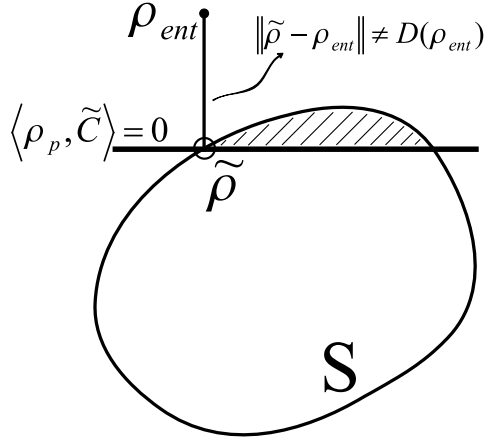}
    \caption{Illustration why $\tilde{C}$ cannot be an entanglement witness if  $\tilde{\rho}$
    is not the nearest separable state. The hatched area is the one were the condition
    $\langle \rho , \tilde{C} \rangle \geq 0 \; \forall \rho \in S$
    is violated.}
    \label{figcheckhs}
\end{figure}

How does it work? Let us start with an entangled state $\rho_{\rm ent}$ and let us call
$\tilde{\rho}$ the guess for the nearest separable state. From previous considerations (Eqs.
(\ref{defc}), (\ref{defplane}) and (\ref{defplanewithc})) we know that the operator
\begin{equation} \label{ctilde}
    \tilde{C} \;=\; \frac{\tilde{\rho} - \rho_{\rm ent} - \left\langle \tilde{\rho} ,
    \tilde{\rho} - \rho_{\rm ent} \right\rangle \mathbbm{1}}{\left\| \tilde{\rho} -
    \rho_{\rm ent} \right\|}
\end{equation}
defines a hyperplane which is orthogonal to $\tilde{\rho} - \rho_{\rm ent}$ and includes
$\tilde{\rho}$. Now we state the following lemma:
\begin{quote}
\textbf{Lemma.} A state $\tilde{\rho}$ is equal to the nearest separable state $\rho_0$
if and only if $\tilde{C}$ is an entanglement witness.
\end{quote}
\emph{Proof.} We already know from Sect.~\ref{sectheorem} that if $\tilde{\rho}$ is the nearest
separable state then the operator $\tilde{C}$ is an entanglement witness. So we need to prove the
opposite: If $\tilde{C}$ is an entanglement witness the state $\tilde{\rho}$ has to be the nearest
separable state $\rho_0$. We prove it indirectly. If $\tilde{\rho}$ is not the nearest separable
state then $\left\| \rho_{\rm ent}-\tilde{\rho} \right\|$ does not give the minimal distance to
$S$; the plane defined by $\langle \rho_p , \tilde{C} \rangle = 0$ is not tangent to $S$ and thus
the existence of `left-hand' separable states $\rho_{\rm sep}\,$ satisfying $\langle \rho_{\rm
sep} , \tilde{C} \rangle < 0\,$ follows. That means $\tilde{C}$ cannot be an entanglement witness
(inequalities (\ref{defentwit}) are not fulfilled), see Fig.~\ref{figcheckhs}.
\\\\
\emph{Remark.} Of course, in general it is not easy to check wether the operator $\tilde{C}$ is an
entanglement witness. However, for some cases (like in Sects.~\ref{secqubitiso},
\ref{secqutritiso} and \ref{secqugeniso}) it is easier to apply the Lemma than using other
procedures to determine the nearest separable state.

If $\tilde{C}$ is indeed an entanglement witness then, because it is tangent to $S$, it is optimal
and can be written as $\tilde{C} = A_{opt}\,$, exactly like Eq. (\ref{entwitmaxviolation}). It is
the operator for which the GBI is maximally violated.

\section{Isotropic Qubit States} \label{secqubitiso}

For illustration we present now examples. In Ref.~\cite{bertlmann02} the 2-qubit Werner
state has been studied -- here we consider the isotropic state in 2 dimensions (acting on
${\cal H}^2 \otimes {\cal H}^2$, it is obtained for $d=2$ in Eqs. (\ref{rhodefiso}),
(\ref{defiso}))
\begin{equation}
    \rho_{\alpha} \;=\; \alpha \left| \phi_+^2 \right\rangle \left\langle \phi_+^2 \right|
    \,+\, \frac{1 - \alpha}{4}\, \mathbbm{1} \;, \qquad - \frac{1}{3} \leq \alpha \leq 1 \,,
\end{equation}
where
\begin{equation}
    \left| \phi_+^2 \right\rangle \;=\; \frac{1}{\sqrt{2}}
    \left( \left| 0 \right\rangle \otimes \left| 0 \right\rangle \,+\,
    \left| 1 \right\rangle \otimes \left| 1 \right\rangle \right)\;.
\end{equation}
In matrix notation in the standard  product basis $\left\{ \left| 0 \right\rangle \otimes
\left|0\right\rangle,
    \left| 0 \right\rangle \otimes \left| 1 \right\rangle,
    \left| 1 \right\rangle \otimes \left| 0 \right\rangle,
    \left| 1 \right\rangle \otimes \left| 1 \right\rangle \right\}$
we get
\begin{equation}
    \rho_{\alpha} \;=\, \left(
    \begin{array}{cccc}
        \frac{\displaystyle 1 + \alpha}{\displaystyle 4} & 0 & 0 &
        \frac{\displaystyle \alpha}{\displaystyle 2} \\
        0 & \frac{\displaystyle 1 - \alpha}{\displaystyle 4} & 0 & 0 \\
        0 & 0 & \frac{\displaystyle 1 - \alpha}{\displaystyle 4} & 0 \\
        \frac{\displaystyle \alpha}{\displaystyle 2} & 0 & 0 &
        \frac{\displaystyle 1 + \alpha}{\displaystyle 4}
    \end{array} \right) \;,
\end{equation}
whereas in terms of the Pauli matrices basis (\ref{statewithpauli}) the state can be
expressed by
\begin{equation} \label{isoqubitrho-sigma}
    \rho_{\alpha} \;=\; \frac{1}{4} \left(\mathbbm{1} + \alpha \, \Sigma \right)\;,
\end{equation}
with the definition
\begin{equation} \label{defbigsigma}
    \Sigma \;:=\; \sigma^x \otimes \sigma^x - \sigma^y \otimes \sigma^y +
    \sigma^z \otimes \sigma^z \;.
\end{equation}
We know that $\rho_{\alpha}$ is (recall Eq.~(\ref{isosepent}))
\begin{equation} \label{isosep}
    \mbox{for}\quad-\frac{1}{3} \leq \alpha \leq \frac{1}{3} \quad\mbox{separable}\,, \qquad
    \mbox{for}\quad\frac{1}{3} < \alpha \leq 1 \quad\mbox{ entangled} \,.
\end{equation}

To compute the Hilbert-Schmidt measure for an entangled isotropic state $\rho_{\alpha}^{\rm ent}$
we need to calculate $D(\rho_{\alpha}^{\rm ent}) = \min_{\rho \in S} \left\| \rho -
\rho_{\alpha}^{\rm ent} \right\|\,$, that is, we need to find the nearest separable state $\rho_0$
to the entangled state in order to obtain $D(\rho_{\alpha}^{\rm ent}) = \left\| \rho_0 -
\rho_{\alpha}^{\rm ent} \right\|\,$. From the separability condition (\ref{isosep}) we see that
the state with $\alpha = 1/3$ lies on the boundary between separable and entangled isotropic
states. Thus our guess  for \emph{all} isotropic entangled qubit states is (and we call it
$\tilde\rho$):
\begin{equation}
    \tilde{\rho} \;=\; \rho_{1/3} \;=\; \frac{1}{4} \left(\mathbbm{1} \,+\,
    \frac{1}{3} \, \Sigma \right).
\end{equation}
Now we have to check that the operator $\tilde{C}$ (\ref{ctilde}) is an entanglement witness (see
Lemma in Sect.~ \ref{secguess}). For this purpose we calculate the expressions
\begin{equation} \label{minus2iso}
    \tilde{\rho} - \rho_{\alpha}^{\rm ent} \;=\; \frac{1}{4} \left( \frac{1}{3} - \alpha \right)
    \Sigma \qquad \mbox{with} \quad \left\| \tilde{\rho} - \rho_{\alpha}^{\rm ent} \right\|
    \;=\; \frac{\sqrt{3}}{2} \left( \alpha - \frac{1}{3} \right) \,,
\end{equation}
(note that $\left\| \Sigma \right\| = 2 \sqrt{3}$) and
\begin{equation} \label{scalar2iso}
    \left\langle \tilde{\rho} , \tilde{\rho} - \rho_{\alpha}^{\rm ent} \right \rangle \;=\;
    \textnormal{Tr}\,\tilde{\rho}(\tilde{\rho} - \rho_{\alpha}^{\rm ent}) \;=\;
    \frac{1}{4} \left( \frac{1}{3} - \alpha \right) \,.
\end{equation}
Then the operator $\tilde{C}$ is explicitly given by
\begin{equation} \label{optent2isotilde}
    \tilde{C} \;=\; \frac{\tilde{\rho} - \rho_{\alpha}^{\rm ent} - \left\langle \tilde{\rho} ,
    \tilde{\rho} - \rho_{\alpha}^{\rm ent} \right\rangle \mathbbm{1}}{\left\| \tilde{\rho} -
    \rho_{\alpha}^{\rm ent} \right\|} \;=\;
    \frac{1}{2\sqrt{3}} \left( \mathbbm{1} - \Sigma \right)\,.
\end{equation}
We examine that $\tilde{C}$ is an entanglement witness, i.e., we check inequalities
(\ref{defentwit}). For the entangled state (where $\alpha > 1/3$) we get
\begin{equation} \label{scalarent2}
    \left\langle \rho_{\alpha}^{\rm ent} , \tilde{C} \right\rangle \;=\;
    \textnormal{Tr} \, \rho_{\alpha}^{\rm ent} \tilde{C} \;=\;
    - \frac{\sqrt{3}}{2} \left( \alpha - \frac{1}{3} \right) \;<\; 0 \,.
\end{equation}
So the first condition is satisfied. The second one, the positivity of $\langle \rho ,
\tilde{C} \rangle$ for all separable states $\rho$ we see in the following way. With
notation (\ref{seppauli}) for $\rho_{\rm sep}$ the scalar product is
\begin{equation} \label{scalarsep2}
    \left\langle \rho_{\rm sep} , \tilde{C} \right\rangle \;=\; \sum_k p_k \ \frac{1}{2\sqrt{3}}
    \left( 1 - n_x^k m_x^k + n_y^k m_y^k - n_z^k m_z^k \right)\,, \qquad
    \left|\vec{n}^k\right| \leq 1, \ \left|\vec{m}^k\right| \leq 1 \;.
\end{equation}
We have to show that
\begin{equation} \label{condpos}
    - n_x^k m_x^k + n_y^k m_y^k - n_z^k m_z^k \;\geq\; -1 \,,
\end{equation}
then the right-hand side of Eq.~(\ref{scalarsep2}) remains always positive. (The convex
sum of positive terms stays positive.) From the property
\begin{equation} \label{propscalar1}
    \left| \vec{n}^k \cdot \vec{m}^k \right| \;\leq\; \left|\vec{n}^k\right|
    \left|\vec{m}^k\right| \;\leq\; 1 \qquad \mbox{or} \qquad
    -1 \;\leq\; \vec{n}^k \cdot \vec{m}^k \;\leq\; 1 \,,
\end{equation}
we find indeed that  Eq. (\ref{condpos}) is satisfied
\begin{equation}
    - n_x^k m_x^k + n_y^k m_y^k - n_z^k m_z^k \;\geq\;
    - n_x^k m_x^k - n_y^k m_y^k - n_z^k m_z^k \;=\;
    - \,\vec{n}^k \cdot \vec{m}^k \;\geq\; -1 \,,
\end{equation}
which completes the proof that $\langle \rho , \tilde{C} \rangle \geq 0 \; \forall \rho
\in S$. So $\tilde{C}$ represents an entanglement witness
\begin{equation}
   A_{opt} \;=\; \tilde{C} \;=\; \frac{1}{2\sqrt{3}} \left( \mathbbm{1} - \Sigma \right)\,,
\end{equation}
and our guess for the nearest separable state was correct, $\tilde{\rho} = \rho_0 \,$.\\

The Hilbert-Schmidt measure for the entangled isotropic state is determined by
Eq.~(\ref{minus2iso}),
\begin{equation} \label{distance2iso3}
    D(\rho_{\alpha}^{\rm ent}) \;=\; \left\| \rho_0 - \rho_{\alpha}^{\rm ent} \right\| \;=\;
    \frac{\sqrt{3}}{2} \left( \alpha - \frac{1}{3} \right)\,.
\end{equation}
It only remains to check the Bertlmann-Narnhofer-Thirring Theorem (\ref{bnttheorem}).
Thus we calculate the maximal violation $B(\rho_{\alpha}^{\rm ent})$
(\ref{maxviolationgbi}) of the GBI. The maximum is attained for the optimal entanglement
witness $A_{opt}$ and the minimum for the nearest separable state $\rho_0\,$. Then Eq.
(\ref{scalarent2}) determines the value of $B(\rho_{\alpha}^{\rm ent})$ (recall that
$\left\langle \rho_0 , A_{opt} \right\rangle = 0$)
\begin{equation}
    B(\rho_{\alpha}^{\rm ent}) \;=\; - \left\langle \rho_{\alpha}^{\rm ent} ,
    A_{opt} \right\rangle \;=\; \frac{\sqrt{3}}{2} \left( \alpha - \frac{1}{3} \right)\,.
\end{equation}
So, indeed $D(\rho_{\alpha}^{\rm ent}) = B(\rho_{\alpha}^{\rm ent})\,$, the Hilbert-Schmidt
measure equals the maximal violation of the GBI.

\section{Isotropic Qutrit States} \label{secqutritiso}

Eqs. (\ref{rhodefiso}) and (\ref{defiso}) define the isotropic qutrit state for $d=3$
\begin{equation}
    \rho_{\alpha} \;=\; \alpha \,\left| \phi_+^3 \right\rangle \left\langle \phi_+^3 \right|
    \,+\, \frac{1 - \alpha}{9} \, \mathbbm{1}\,, \qquad - \frac{1}{8} \leq \alpha \leq 1 \,,
\end{equation}
where
\begin{equation}
    \left| \phi_+^3 \right\rangle \;=\;
    \frac{1}{\sqrt{3}} \,\Big( \left| 0 \right\rangle \otimes \left| 0 \right\rangle +
    \left| 1 \right\rangle \otimes \left| 1 \right\rangle +
    \left| 2 \right\rangle \otimes \left| 2 \right\rangle \Big) \,.
\end{equation}
In matrix notation in the standard product basis
\begin{displaymath}
\left\{ \left| 0 \right\rangle \otimes \left| 0 \right\rangle,
    \left| 0 \right\rangle \otimes \left| 1 \right\rangle,
    \left| 0 \right\rangle \otimes \left| 2 \right\rangle,
    \left| 1 \right\rangle \otimes \left| 0 \right\rangle,
    \left| 1 \right\rangle \otimes \left| 1 \right\rangle,
    \left| 1 \right\rangle \otimes \left| 2 \right\rangle,
    \left| 2 \right\rangle \otimes \left| 0 \right\rangle,
    \left| 2 \right\rangle \otimes \left| 1 \right\rangle,
    \left| 2 \right\rangle \otimes \left| 2 \right\rangle \right\}
\end{displaymath}
we have
\begin{equation}
    \rho_{\alpha} \;=\; \left(
    \begin{array}{ccccccccc}
        \frac{\displaystyle 1 + 2\alpha}{\displaystyle 9} & 0 & 0 & 0 &
        \frac{\displaystyle \alpha}{\displaystyle 3} & 0 & 0 & 0 &
         \frac{\displaystyle \alpha}{\displaystyle 3} \\
        0 & \frac{\displaystyle 1 - \alpha}{\displaystyle 9} & 0 & 0 & 0 & 0 & 0 & 0 & 0 \\
        0 & 0 & \frac{\displaystyle 1 - \alpha}{\displaystyle 9} & 0 & 0 & 0 & 0 & 0 & 0 \\
        0 & 0 & 0 & \frac{\displaystyle 1 - \alpha}{\displaystyle 9} & 0 & 0 & 0 & 0 & 0 \\
        \frac{\displaystyle \alpha}{\displaystyle 3} & 0 & 0 & 0 &
        \frac{\displaystyle 1 + 2\alpha}{\displaystyle 9} & 0 & 0 & 0 &
         \frac{\displaystyle \alpha}{\displaystyle 3} \\
        0 & 0 & 0 & 0 & 0 & \frac{\displaystyle 1 - \alpha}{\displaystyle 9} & 0 & 0 & 0 \\
        0 & 0 & 0 & 0 & 0 & 0 & \frac{\displaystyle 1 - \alpha}{\displaystyle 9} & 0 & 0 \\
        0 & 0 & 0 & 0 & 0 & 0 & 0 & \frac{\displaystyle 1 - \alpha}{\displaystyle 9} & 0 \\
        \frac{\displaystyle \alpha}{\displaystyle 3} & 0 & 0 & 0 &
        \frac{\displaystyle \alpha}{\displaystyle 3} & 0 & 0 & 0 &
         \frac{\displaystyle 1 + 2\alpha}{\displaystyle 9}
    \end{array} \right)\;.
\end{equation}
In the Gell-Mann matrices representation (\ref{statewithgellmann}) the state
$\rho_{\alpha}$ can be expressed by (see also Ref. \cite{caves00})
\begin{equation} \label{iso3dshortgellmann}
    \rho_{\alpha} \;=\; \frac{1}{9} \left(\mathbbm{1} + \frac{3 \alpha}{2} \Lambda \right)\,,
\end{equation}
with the definition
\begin{equation}
    \Lambda \;:=\; \lambda^1 \otimes \lambda^1 - \lambda^2 \otimes \lambda^2 +
    \lambda^3 \otimes \lambda^3 + \lambda^4 \otimes \lambda^4 -
    \lambda^5 \otimes \lambda^5 + \lambda^6 \otimes \lambda^6 -
    \lambda^7 \otimes \lambda^7 + \lambda^8 \otimes \lambda^8 \,.
\end{equation}
From Eq.~(\ref{isosepent}) we know that
\begin{equation} \label{iso3dsepent}
    \begin{array}{ccc}
        -\frac{\displaystyle 1}{\displaystyle 8} \leq \alpha \leq
        \frac{\displaystyle 1}{\displaystyle 4} & \quad\Rightarrow\quad &
        \rho_{\alpha} \;\mbox{ separable} \,, \\[2ex]
        \frac{\displaystyle 1}{\displaystyle 4} < \alpha \leq 1 & \quad\Rightarrow\quad &
        \rho_{\alpha} \;\mbox{ entangled} \,.
    \end{array}
\end{equation}

By the same argument as in the qubit case we guess the nearest separable state to the
state (\ref{iso3dshortgellmann})
\begin{equation}
    \tilde{\rho} \;=\; \rho_{1/4} \;=\; \frac{1}{9}
    \left(\mathbbm{1} + \frac{3}{8} \Lambda \right) \,.
\end{equation}
Again, to check our guess we examine that the operator $\tilde{C}$ (\ref{ctilde}) is an
entanglement witness. We need the following expressions
\begin{equation} \label{minus3iso}
    \tilde{\rho} - \rho_{\alpha}^{\rm ent} \;=\; \frac{1}{6} \left( \frac{1}{4} - \alpha \right)
    \Lambda \qquad \mbox{with} \quad \left\| \tilde{\rho} - \rho_{\alpha}^{\rm ent} \right\|
    \;=\; \frac{2 \sqrt{2}}{3} \left( \alpha - \frac{1}{4} \right) \,,
\end{equation}
(where $\left\| \Lambda \right\| = 4 \sqrt{2}$) and
\begin{equation} \label{scalar3iso}
    \left\langle \tilde{\rho} , \tilde{\rho} - \rho_{\alpha}^{\rm ent} \right \rangle \;=\;
    \textnormal{Tr}\,\tilde{\rho}(\tilde{\rho} - \rho_{\alpha}^{\rm ent}) \;=\;
    \frac{2}{9} \left( \frac{1}{4} - \alpha \right) \,.
\end{equation}
Then $\tilde{C}$ (\ref{ctilde}) is explicitly given by
\begin{equation} \label{optent3isotilde}
    \tilde{C} \;=\; \frac{1}{3\sqrt{2}} \left( \mathbbm{1} - \frac{3}{4} \Lambda \right)\,.
\end{equation}

Now let us check the entanglement witness conditions (\ref{defentwit}) for $ \tilde{C}$
\begin{equation} \label{scalarent3}
    \left\langle \rho_{\alpha}^{\rm ent} , \tilde{C} \right\rangle \;=\;
    \textnormal{Tr} \,\rho_{\alpha}^{\rm ent} \tilde{C} \;=\;
    - \frac{2 \sqrt{2}}{3} \left( \alpha - \frac{1}{4} \right) \;<\; 0 \,.
\end{equation}
So the first condition is satisfied since $\alpha > 1/4$; for the second one we obtain
\begin{eqnarray} \label{scalarsep3}
    \left\langle \rho_{\rm sep} , \tilde{C} \right\rangle \;=\;
    & \sum_k & p_k \, \frac{1}{3\sqrt{2}} \,\big{(} 1 - n_1^k m_1^k +
    n_2^k m_2^k - n_3^k m_3^k - n_4^k m_4^k  +  n_5^k m_5^k \nonumber\\
    & - & n_6^k m_6^k \,+\, n_7^k m_7^k \,-\, n_8^k m_8^k \big{)}\,, \qquad\quad
    \left|\vec{n}^k\right| \leq 1, \ \left|\vec{m}^k\right| \leq 1 \;.
\end{eqnarray}
Since the inequalities (\ref{propscalar1}) apply here as well we have
\begin{equation}
    - n_1^k m_1^k + n_2^k m_2^k - n_3^k m_3^k - n_4^k m_4^k + n_5^k m_5^k - n_6^k m_6^k +
    n_7^k m_7^k - n_8^k m_8^k \;\geq\; - \vec{n}^k \cdot \vec{m}^k \;\geq\; -1 \,,
\end{equation}
so that $\langle \rho_{\rm sep} , \tilde{C} \rangle \geq 0 \,$. Indeed, $\tilde{C}$
represents an entanglement witness and we identify
\begin{equation}
    A_{opt} \;=\; \tilde{C} \;=\; \frac{1}{3\sqrt{2}}
    \left( \mathbbm{1} - \frac{3}{4}\Lambda \right) \qquad
    \mbox{and} \qquad \tilde{\rho} \;=\; \rho_0 \,.
\end{equation}

With Eq.~(\ref{minus3iso}) the Hilbert-Schmidt measure is
\begin{equation}
    D(\rho_{\alpha}^{\rm ent}) \;=\; \left\| \rho_0 - \rho_{\alpha}^{\rm ent} \right\| \;=\;
    \frac{2\sqrt{2}}{3} \left( \alpha - \frac{1}{4} \right) \,,
\end{equation}
and by the same argumentation as for qubits the maximal violation $B(\rho_{\alpha}^{\rm ent})$
(\ref{maxviolationgbi}) of the GBI is determined by Eq. (\ref{scalarent3})
\begin{equation}
    B(\rho_{\alpha}^{\rm ent}) \;=\; - \left\langle \rho_{\alpha}^{\rm ent} , A_{opt} \right\rangle
    \;=\; \frac{2\sqrt{2}}{3} \left( \alpha - \frac{1}{4} \right) \,.
\end{equation}
So again, $D(\rho_{\alpha}^{\rm ent}) = B(\rho_{\alpha}^{\rm ent})\,$, we see that the
Bertlmann-Narnhofer-Thirring Theorem is satisfied.

\section{Isotropic States in Higher Dimensions} \label{secqugeniso}

Finally, we want to show how we can generalize our isotropic qubit and qutrit results to
arbitrary dimensions. A general state on ${\cal H}^d$ can be written in a matrix basis
$\left\{ \mathbbm{1}, \gamma^1, \ \ldots \ ,\gamma^{d^2-1} \right\}$ by
\begin{equation} \label{omega-d}
    \omega \;=\; \frac{1}{d} \left( \mathbbm{1} +
    \sqrt{\frac{d(d-1)}{2}} \,n_i\,\gamma^i \right)\,,
    \qquad \sum_i n_i^2 =: \left| \vec{n} \right|^2 \leq 1 \,.
\end{equation}
We have included the factor $\sqrt{\frac{d(d-1)}{2}}$ for the
correct normalization  and the matrices $\gamma^i$ have the
properties
\begin{equation} \label{gamma-properties}
    \textnormal{Tr}\,\gamma^i \;=\; 0\,, \quad
    \textnormal{Tr}\,\gamma^i \gamma^j \;=\; 2 \,\delta^{ij} \,.
\end{equation}
Considering the tensor product space ${\cal H}^d \otimes {\cal H}^d$ the notation of
separable states is a straight forward extension to Eqs. (\ref{seppauli}) and
(\ref{sepgellmann})
\begin{eqnarray} \label{sepmatrixgeneral}
    \rho_{\rm sep} \;=\; &\sum_k& p_k \, \frac{\textstyle 1}{\textstyle d^2}\,
    \bigg{(} \mathbbm{1} \otimes \mathbbm{1} \;+\;
    \sqrt{\frac{d(d-1)}{2}}\;n_i^k\;\gamma^i \otimes \mathbbm{1}\nonumber\\
    \,&+&\, \sqrt{\frac{d(d-1)}{2}}\; m_i^k\;\mathbbm{1} \otimes \gamma^i \;+\;
    \frac{d(d-1)}{2} \;n_i^k m_j^k \;\gamma^i \otimes \gamma^j \bigg{)} \;.
\end{eqnarray}
A $d\times d$-dimensional isotropic state -- as a generalization of the isotropic qubit state
(\ref{isoqubitrho-sigma}) and qutrit state (\ref{iso3dshortgellmann}) -- we express as
\begin{equation} \label{isogamma}
    \rho_{\alpha} \;=\; \frac{\textstyle 1}{\textstyle d^2} \left( \mathbbm{1} \,+\,
    \frac{\textstyle d}{\textstyle 2} \,\alpha\, \Gamma \right)\,, \qquad
    -\frac{1}{d^2-1} \leq \alpha \leq 1 \,,
\end{equation}
where we define
\begin{equation} \label{Gamma}
    \Gamma \;:=\; \sum_{i=1}^{d^2-1} \,c_i \, \gamma^i \otimes \gamma^i \,, \quad
    c_i = \pm 1 \;.
\end{equation}
The factor $\frac{d}{2}$ in Eq.~(\ref{isogamma}) is due to normalization. The splitting
of $\rho_{\alpha}$ into entangled and separable states is given by Eq. (\ref{isosepent}).

There is strong evidence that expression (\ref{isogamma}) with definition (\ref{Gamma})
coincides with the isotropic state definition (\ref{rhodefiso}), (\ref{defiso}), which we
introduced in the beginning, for all dimensions $d\times d$. That means, there exist $d^2
- 1$ matrices $\gamma^i$ with properties (\ref{gamma-properties}), which form a basis
together with the identity $\mathbbm{1}$ for all $d^2 \times d^2$ matrices. They describe
the quantum state in the isotropic way (\ref{isogamma}), (\ref{Gamma}) and can be
expressed as linear-combinations of density matrix elements in the standard basis
notation.\\

In this way a generalization of our previous results is possible and can be obtained by
calculations very similar to the ones for qubits and qutrits (see Sect.~\ref{secqubitiso}
and Sect.~\ref{secqutritiso}). In particular, using the same notations as before, we find
the following expressions for the nearest separable state $\rho_0$, the Hilbert-Schmidt
measure $D(\rho_{\alpha}^{\rm ent})$ and the optimal entanglement witness $A_{opt}\,$:
\begin{equation}
    \rho_0 \;=\; \rho_{\frac{1}{d+1}} \;=\; \frac{1}{d^2} \left( \mathbbm{1} \,+\,
    \frac{d}{2(d+1)} \;\Gamma \right)\,,
\end{equation}
\begin{equation}
    D(\rho_{\alpha}^{\rm ent}) \;=\; \left\| \rho_0 - \rho_{\alpha}^{\rm ent} \right\| \;=\;
    \frac{\sqrt{d^2-1}}{d} \left( \alpha \,-\, \frac{1}{d+1} \right)\,,
\end{equation}
\begin{equation}
    A_{opt} \;=\; \frac{d-1}{d\sqrt{d^2-1}}
    \left( \mathbbm{1} \,-\, \frac{d}{2(d-1)} \;\Gamma \right)\,.
\end{equation}
The maximal violation of the GBI gives
\begin{equation}
    B(\rho_{\alpha}^{\rm ent}) \;=\;
    - \left\langle \rho_{\alpha}^{\rm ent} , A_{opt} \right\rangle \;=\;
    \frac{\sqrt{d^2-1}}{d} \left( \alpha \,-\, \frac{1}{d+1} \right)\,,
\end{equation}
thus we see that again $D(\rho_{\alpha}^{\rm ent}) = B(\rho_{\alpha}^{\rm ent})\,$ and
Theorem (\ref{bnttheorem}) is satisfied.\\

\emph{Remark.} For the limit of infinite dimensions, $d\rightarrow\infty\,$, the distance
or the maximal violation of GBI approaches the parameter $\alpha\,$, that means, the
region where the isotropic state is separable shrinks to zero (see in this connection
Refs.~\cite{zyczkowski98, zyczkowski99}).

\section{Conclusions}

In this paper we enlighten the connection between the Hilbert-Schmidt measure of
entanglement and an optimal entanglement witness. This connection is viewed via the
Bertlmann-Narnhofer-Thirring Theorem (\ref{bnttheorem}) which states that the
Hilbert-Schmidt measure equals the maximal violation of a generalized Bell inequality.
This inequality detects entanglement versus separability and not like the original Bell
inequality non-locality versus locality. Furthermore, we present a method how to guess
the nearest separable state to a given entangled state. We illustrate the general results
with the examples of isotropic qubit and qutrit states and show a possible generalization
of the method for isotropic states of higher dimensions.

However, we remark that in general for non-isotropic states the situation might turn out
to be rather different. The reason is that our method for constructing an optimal
entanglement witness, Eq.~(\ref{entwitmaxviolation}), involves the nearest separable
state $\rho_0$ to a given entangled one, which in general might turn out to be a
difficult task. But in some cases, like in the case of isotropic states, the Lemma
presented in the article will be helpful to use.\\

\begin{acknowledgments}

We would like to thank Heide Narnhofer and Walter Thirring for helpful comments. This research has
been supported by the EU project EURIDICE HPRN-CT-2002-00311.

\end{acknowledgments}

\bibliography{witnessref}

\begin{thebibliography}{31}
\expandafter\ifx\csname natexlab\endcsname\relax\def\natexlab#1{#1}\fi
\expandafter\ifx\csname bibnamefont\endcsname\relax
  \def\bibnamefont#1{#1}\fi
\expandafter\ifx\csname bibfnamefont\endcsname\relax
  \def\bibfnamefont#1{#1}\fi
\expandafter\ifx\csname citenamefont\endcsname\relax
  \def\citenamefont#1{#1}\fi
\expandafter\ifx\csname url\endcsname\relax
  \def\url#1{\texttt{#1}}\fi
\expandafter\ifx\csname urlprefix\endcsname\relax\def\urlprefix{URL }\fi
\providecommand{\bibinfo}[2]{#2}
\providecommand{\eprint}[2][]{\url{#2}}

\bibitem[{\citenamefont{Schroedinger}(1935)}]{schroedinger35}
\bibinfo{author}{\bibfnamefont{E.}~\bibnamefont{Schr\"odinger}},
  \bibinfo{journal}{Naturwissenschaften} \textbf{\bibinfo{volume}{23}},
  \bibinfo{pages}{807} (\bibinfo{year}{1935});
  \textbf{\bibinfo{volume}{23}},
  \bibinfo{pages}{823} (\bibinfo{year}{1935});
  \textbf{\bibinfo{volume}{23}},
  \bibinfo{pages}{844} (\bibinfo{year}{1935}).

\bibitem[{\citenamefont{Einstein et~al.}(1935)\citenamefont{Einstein, Podolsky,
  and Rosen}}]{einstein35}
\bibinfo{author}{\bibfnamefont{A.}~\bibnamefont{Einstein}},
  \bibinfo{author}{\bibfnamefont{B.}~\bibnamefont{Podolsky}}, \bibnamefont{and}
  \bibinfo{author}{\bibfnamefont{N.}~\bibnamefont{Rosen}},
  \bibinfo{journal}{Phys. Rev.} \textbf{\bibinfo{volume}{47}},
  \bibinfo{pages}{777} (\bibinfo{year}{1935}).

\bibitem[{\citenamefont{Bertlmann and Zeilinger}(2002)}]{bertlmann02a}
\bibinfo{editor}{\bibfnamefont{R.~A.} \bibnamefont{Bertlmann}}
  \bibnamefont{and}
  \bibinfo{editor}{\bibfnamefont{A.}~\bibnamefont{Zeilinger}}, eds.,
  \emph{\bibinfo{title}{Quantum \emph{[Un]}speakables, from Bell to Quantum Information}}
  (\bibinfo{publisher}{Springer, Berlin Heidelberg New York}, \bibinfo{year}{2002}).

\bibitem[{\citenamefont{Brukner et~al.}(2002)\citenamefont{Brukner, \.Zukowski,
  and Zeilinger}}]{brukner02}
\bibinfo{author}{\bibfnamefont{\v C.}~\bibnamefont{Brukner}},
  \bibinfo{author}{\bibfnamefont{M.}~\bibnamefont{\.Zukowski}}, \bibnamefont{and}
  \bibinfo{author}{\bibfnamefont{A.}~\bibnamefont{Zeilinger}},
  \bibinfo{journal}{Phys. Rev. Lett.} \textbf{\bibinfo{volume}{89}},
  \bibinfo{pages}{197901} (\bibinfo{year}{2002}).

\bibitem[{\citenamefont{Bru{\ss}}(2002)}]{bruss02}
\bibinfo{author}{\bibfnamefont{D.}~\bibnamefont{Bru{\ss}}},
  \bibinfo{journal}{J. Math. Phys.} \textbf{\bibinfo{volume}{43}},
  \bibinfo{pages}{4237} (\bibinfo{year}{2002}).

\bibitem[{\citenamefont{Horodecki et~al.}(2001)\citenamefont{Horodecki,
  Horodecki, and Horodecki}}]{horodecki01}
\bibinfo{author}{\bibfnamefont{M.}~\bibnamefont{Horodecki}},
  \bibinfo{author}{\bibfnamefont{P.}~\bibnamefont{Horodecki}},
  \bibnamefont{and}
  \bibinfo{author}{\bibfnamefont{R.}~\bibnamefont{Horodecki}}, in
  \emph{\bibinfo{booktitle}{Quantum Information}},
  \bibinfo{editor}{\bibfnamefont{G.~Alber} \bibnamefont{et~al.}}, eds.,
  \emph{\bibinfo{series}{Springer Tracts in Modern Physics}} \bibinfo{volume}{173}
  (\bibinfo{publisher}{Springer Verlag Berlin}, \bibinfo{year}{2001}) p. \bibinfo{pages}{151}.

\bibitem[{\citenamefont{Peres}(1996)}]{peres96}
\bibinfo{author}{\bibfnamefont{A.}~\bibnamefont{Peres}},
  \bibinfo{journal}{Phys. Rev. Lett.} \textbf{\bibinfo{volume}{77}},
  \bibinfo{pages}{1413} (\bibinfo{year}{1996}).

\bibitem[{\citenamefont{Horodecki et~al.}(1996)\citenamefont{Horodecki,
  Horodecki, and Horodecki}}]{horodecki96}
\bibinfo{author}{\bibfnamefont{M.}~\bibnamefont{Horodecki}},
  \bibinfo{author}{\bibfnamefont{P.}~\bibnamefont{Horodecki}},
  \bibnamefont{and}
  \bibinfo{author}{\bibfnamefont{R.}~\bibnamefont{Horodecki}},
  \bibinfo{journal}{Phys. Lett. A} \textbf{\bibinfo{volume}{223}},
  \bibinfo{pages}{1} (\bibinfo{year}{1996}).

\bibitem[{\citenamefont{Bennett et~al.}(1996)\citenamefont{Bennett, DiVincenzo,
  Smolin, and Wootters}}]{bennett96}
\bibinfo{author}{\bibfnamefont{C.~H.} \bibnamefont{Bennett}},
  \bibinfo{author}{\bibfnamefont{D.~P.} \bibnamefont{DiVincenzo}},
  \bibinfo{author}{\bibfnamefont{J.~A.} \bibnamefont{Smolin}},
  \bibnamefont{and} \bibinfo{author}{\bibfnamefont{W.~K.}
  \bibnamefont{Wootters}}, \bibinfo{journal}{Phys. Rev. A}
  \textbf{\bibinfo{volume}{54}}, \bibinfo{pages}{3824} (\bibinfo{year}{1996}).

\bibitem[{\citenamefont{Hill and Wootters}(1997)}]{hill97}
\bibinfo{author}{\bibfnamefont{S.}~\bibnamefont{Hill}} \bibnamefont{and}
  \bibinfo{author}{\bibfnamefont{W.~K.} \bibnamefont{Wootters}},
  \bibinfo{journal}{Phys. Rev. Lett.} \textbf{\bibinfo{volume}{78}},
  \bibinfo{pages}{5022} (\bibinfo{year}{1997}).

\bibitem[{\citenamefont{Wootters}(1998)}]{wootters98}
\bibinfo{author}{\bibfnamefont{W.~K.} \bibnamefont{Wootters}},
  \bibinfo{journal}{Phys. Rev. Lett.} \textbf{\bibinfo{volume}{80}},
  \bibinfo{pages}{2245} (\bibinfo{year}{1998}).

\bibitem[{\citenamefont{Vedral et~al.}(1997)\citenamefont{Vedral, Plenio,
  Rippin, and Knight}}]{vedral97}
\bibinfo{author}{\bibfnamefont{V.}~\bibnamefont{Vedral}},
  \bibinfo{author}{\bibfnamefont{M.~B.} \bibnamefont{Plenio}},
  \bibinfo{author}{\bibfnamefont{M.~A.} \bibnamefont{Rippin}},
  \bibnamefont{and} \bibinfo{author}{\bibfnamefont{P.~L.}
  \bibnamefont{Knight}}, \bibinfo{journal}{Phys. Rev. Lett.}
  \textbf{\bibinfo{volume}{78}}, \bibinfo{pages}{2275} (\bibinfo{year}{1997}).

\bibitem[{\citenamefont{Vedral and Plenio}(1998)}]{vedral98}
\bibinfo{author}{\bibfnamefont{V.}~\bibnamefont{Vedral}} \bibnamefont{and}
  \bibinfo{author}{\bibfnamefont{M.~B.} \bibnamefont{Plenio}},
  \bibinfo{journal}{Phys. Rev. A} \textbf{\bibinfo{volume}{57}},
  \bibinfo{pages}{1619} (\bibinfo{year}{1998}).

\bibitem[{\citenamefont{Witte and Trucks}(1999)}]{witte99}
\bibinfo{author}{\bibfnamefont{C.}~\bibnamefont{Witte}} \bibnamefont{and}
  \bibinfo{author}{\bibfnamefont{M.}~\bibnamefont{Trucks}},
  \bibinfo{journal}{Phys. Lett. A} \textbf{\bibinfo{volume}{257}},
  \bibinfo{pages}{14} (\bibinfo{year}{1999}).

\bibitem[{\citenamefont{Ozawa}(2000)}]{ozawa00}
\bibinfo{author}{\bibfnamefont{M.}~\bibnamefont{Ozawa}},
  \bibinfo{journal}{Phys. Lett. A} \textbf{\bibinfo{volume}{268}},
  \bibinfo{pages}{158} (\bibinfo{year}{2000}).

\bibitem[{\citenamefont{Bertlmann et~al.}(2002)\citenamefont{Bertlmann,
  Narnhofer, and Thirring}}]{bertlmann02}
\bibinfo{author}{\bibfnamefont{R.~A.} \bibnamefont{Bertlmann}},
  \bibinfo{author}{\bibfnamefont{H.}~\bibnamefont{Narnhofer}},
  \bibnamefont{and} \bibinfo{author}{\bibfnamefont{W.}~\bibnamefont{Thirring}},
  \bibinfo{journal}{Phys. Rev. A} \textbf{\bibinfo{volume}{66}},
  \bibinfo{pages}{032319} (\bibinfo{year}{2002}).

\bibitem[{\citenamefont{Horodecki and Horodecki}(1999)}]{horodecki99}
\bibinfo{author}{\bibfnamefont{M.}~\bibnamefont{Horodecki}} \bibnamefont{and}
  \bibinfo{author}{\bibfnamefont{P.}~\bibnamefont{Horodecki}},
  \bibinfo{journal}{Phys. Rev. A} \textbf{\bibinfo{volume}{59}},
  \bibinfo{pages}{4206} (\bibinfo{year}{1999}).

\bibitem[{\citenamefont{Rains}(1999)}]{rains99}
\bibinfo{author}{\bibfnamefont{E.~M.} \bibnamefont{Rains}},
  \bibinfo{journal}{Phys. Rev. A} \textbf{\bibinfo{volume}{60}},
  \bibinfo{pages}{179} (\bibinfo{year}{1999}).

\bibitem[{\citenamefont{Henley and Thirring}(1962)}]{henley62}
\bibinfo{author}{\bibfnamefont{E.~M.} \bibnamefont{Henley}} \bibnamefont{and}
  \bibinfo{author}{\bibfnamefont{W.}~\bibnamefont{Thirring}},
  \emph{\bibinfo{title}{Elementary Quantum Field Theory}}
  (\bibinfo{publisher}{McGraw Hill, New York}, \bibinfo{year}{1962}).

\bibitem[{\citenamefont{Terhal}(2000)}]{terhal00}
\bibinfo{author}{\bibfnamefont{B.~M.} \bibnamefont{Terhal}},
  \bibinfo{journal}{Phys. Lett. A} \textbf{\bibinfo{volume}{271}},
  \bibinfo{pages}{319} (\bibinfo{year}{2000}).

\bibitem[{\citenamefont{Terhal}(2002)}]{terhal02}
\bibinfo{author}{\bibfnamefont{B.~M.} \bibnamefont{Terhal}},
  \bibinfo{journal}{Journal of Theoretical Computer Science}
  \textbf{\bibinfo{volume}{287}}, \bibinfo{pages}{313} (\bibinfo{year}{2002}).

\bibitem[{\citenamefont{Bertlmann and Hiesmayr}(2001)}]{bertlmann01}
\bibinfo{author}{\bibfnamefont{R.~A.} \bibnamefont{Bertlmann}}
  \bibnamefont{and} \bibinfo{author}{\bibfnamefont{B.~C.}
  \bibnamefont{Hiesmayr}}, \bibinfo{journal}{Phys. Rev. A}
  \textbf{\bibinfo{volume}{63}}, \bibinfo{pages}{062112}
  (\bibinfo{year}{2001}).

\bibitem[{\citenamefont{Clauser et~al.}(1969)\citenamefont{Clauser, Horne,
  Shimony, and Holt}}]{clauser69}
\bibinfo{author}{\bibfnamefont{J.~F.} \bibnamefont{Clauser}},
  \bibinfo{author}{\bibfnamefont{M.~A.} \bibnamefont{Horne}},
  \bibinfo{author}{\bibfnamefont{A.}~\bibnamefont{Shimony}}, \bibnamefont{and}
  \bibinfo{author}{\bibfnamefont{R.~A.} \bibnamefont{Holt}},
  \bibinfo{journal}{Phys. Rev. Lett.} \textbf{\bibinfo{volume}{23}},
  \bibinfo{pages}{880} (\bibinfo{year}{1969}).

\bibitem[{\citenamefont{Werner}(1989)}]{werner89}
\bibinfo{author}{\bibfnamefont{R.~F.} \bibnamefont{Werner}},
  \bibinfo{journal}{Phys. Rev. A} \textbf{\bibinfo{volume}{40}},
  \bibinfo{pages}{4277} (\bibinfo{year}{1989}).
 
\bibitem[{\citenamefont{Brand\~ao}(1989)}]{Brandao}
\bibinfo{author}{\bibfnamefont{F.~G.~S.~L.} \bibnamefont{Brand\~ao}},
  \bibinfo{journal}{Phys. Rev. A} \textbf{\bibinfo{volume}{72}},
  \bibinfo{pages}{022310} (\bibinfo{year}{2005}).

\bibitem[{\citenamefont{Arvind, K.~S.~Mallesh, and N.~Mukunda}(1997)}]{arvind97}
\bibinfo{author}{\bibfnamefont{ } \bibnamefont{Arvind}},
  \bibinfo{author}{\bibfnamefont{K.~S.}~\bibnamefont{Mallesh}}, \bibnamefont{and}
  \bibinfo{author}{\bibfnamefont{N.} \bibnamefont{Mukunda}},
  \bibinfo{journal}{J. Phys. A: Math. Gen.} \textbf{\bibinfo{volume}{30}},
  \bibinfo{pages}{2417} (\bibinfo{year}{1997}).

\bibitem[{\citenamefont{Caves and Milburn}(2000)}]{caves00}
\bibinfo{author}{\bibfnamefont{C.~M.} \bibnamefont{Caves}} \bibnamefont{and}
  \bibinfo{author}{\bibfnamefont{G.~J.} \bibnamefont{Milburn}},
  \bibinfo{journal}{Optics Communications} \textbf{\bibinfo{volume}{179}},
  \bibinfo{pages}{439} (\bibinfo{year}{2000}).

\bibitem[{\citenamefont{Pittenger and Rubin}(2003)}]{pittenger03}
\bibinfo{author}{\bibfnamefont{A.~O.} \bibnamefont{Pittenger}}
  \bibnamefont{and} \bibinfo{author}{\bibfnamefont{M.~H.} \bibnamefont{Rubin}},
  \bibinfo{journal}{Phys. Rev. A} \textbf{\bibinfo{volume}{67}},
  \bibinfo{pages}{012327} (\bibinfo{year}{2003}).

\bibitem[{\citenamefont{Lewenstein et~al.}(2000)\citenamefont{Lewenstein,
  Kraus, Cirac, and Horodecki}}]{lewenstein00}
\bibinfo{author}{\bibfnamefont{M.}~\bibnamefont{Lewenstein}},
  \bibinfo{author}{\bibfnamefont{B.}~\bibnamefont{Kraus}},
  \bibinfo{author}{\bibfnamefont{J.~I.} \bibnamefont{Cirac}}, \bibnamefont{and}
  \bibinfo{author}{\bibfnamefont{P.}~\bibnamefont{Horodecki}},
  \bibinfo{journal}{Phys. Rev. A} \textbf{\bibinfo{volume}{62}},
  \bibinfo{pages}{052310} (\bibinfo{year}{2000}).

\bibitem[{\citenamefont{Guehne et~al.}(2003)\citenamefont{Guehne, Hyllus,
  Bru{\ss}, Ekert, Macchiavello, and Sanpera}}]{guehne03}
\bibinfo{author}{\bibfnamefont{O.}~\bibnamefont{Guehne}},
  \bibinfo{author}{\bibfnamefont{P.}~\bibnamefont{Hyllus}},
  \bibinfo{author}{\bibfnamefont{D.}~\bibnamefont{Bru{\ss}}},
  \bibinfo{author}{\bibfnamefont{A.} \bibnamefont{Ekert}},
  \bibinfo{author}{\bibfnamefont{C.}~\bibnamefont{Macchiavello}},
  \bibnamefont{and} \bibinfo{author}{\bibfnamefont{A.}~\bibnamefont{Sanpera}},
  \bibinfo{journal}{J. Mod. Opt.} \textbf{\bibinfo{volume}{50}},
  \bibinfo{pages}{1079} (\bibinfo{year}{2003}).

\bibitem[{\citenamefont{Ioannou et~al.}(2004)\citenamefont{Ioannou, Travaglione,
  Cheung, Ekert}}]{ioannou04}
\bibinfo{author}{\bibfnamefont{L.~M.}~\bibnamefont{Ioannou}},
  \bibinfo{author}{\bibfnamefont{B.~C.}~\bibnamefont{Travaglione}},
  \bibinfo{author}{\bibfnamefont{D.}~\bibnamefont{Cheung}},
  \bibnamefont{and} \bibinfo{author}{\bibfnamefont{A.} \bibnamefont{Ekert}},
  \bibinfo{journal}{Phys. Rev. A} \textbf{\bibinfo{volume}{70}},
  \bibinfo{pages}{060303} \bibinfo{year}{2004}.

\bibitem[{\citenamefont{Verstraete et~al.}(2002)\citenamefont{Verstraete,
  Dehaene, and Moor}}]{verstraete01}
\bibinfo{author}{\bibfnamefont{F.}~\bibnamefont{Verstraete}},
  \bibinfo{author}{\bibfnamefont{J.}~\bibnamefont{Dehaene}}, \bibnamefont{and}
  \bibinfo{author}{\bibfnamefont{B.~D.} \bibnamefont{Moor}},
  \bibinfo{journal}{J. Mod. Opt.} \textbf{\bibinfo{volume}{49}},
  \bibinfo{pages}{1277} (\bibinfo{year}{2002}).

\bibitem[{\citenamefont{Zyczkowski et~al.}(1998)\citenamefont{\.Zyczkowski,
  Horodecki, Sanpera, and Lewenstein}}]{zyczkowski98}
\bibinfo{author}{\bibfnamefont{K.}~\bibnamefont{\.Zyczkowski}},
  \bibinfo{author}{\bibfnamefont{P.}~\bibnamefont{Horodecki}},
  \bibinfo{author}{\bibfnamefont{A.}~\bibnamefont{Sanpera}}, \bibnamefont{and}
  \bibinfo{author}{\bibfnamefont{M.}~\bibnamefont{Lewenstein}},
  \bibinfo{journal}{Phys. Rev. A} \textbf{\bibinfo{volume}{58}},
  \bibinfo{pages}{883} (\bibinfo{year}{1998}).

\bibitem[{\citenamefont{Zyczkowski}(1998)}]{zyczkowski99}
\bibinfo{author}{\bibfnamefont{K.}~\bibnamefont{\.Zyczkowski}},
  \bibinfo{journal}{Phys. Rev. A} \textbf{\bibinfo{volume}{60}},
  \bibinfo{pages}{3496} (\bibinfo{year}{1998}).


\end{thebibliography}

\end{document}